\documentclass[aps,prapplied,reprint]{revtex4-2}
\usepackage{graphicx}
\usepackage{babel}
\usepackage{amsmath,bbm}
\usepackage{amssymb}
\usepackage{xcolor}

\usepackage{mathtools}

\newcommand{\Tr}{\textrm{Tr}}

\newcommand{\ket}[2][]{{|#2\rangle_{#1}}}
\newcommand{\bra}[2][]{{}_{#1}\langle #2|}

\newcommand{\mj}[1]{{\color{black}#1}}

\begin{document}

\title{Photon Efficiency of High-Dimensional Quantum Key Distribution}

\author{Vera Uzunova}%\thanks{e-mail: m.jarzyna@cent.uw.edu.pl}
\affiliation{Centre for Quantum Optical Technologies, Centre of New Technologies, University of Warsaw, Banacha 2c, 02-097 Warszawa, Poland} \affiliation{ Institute of Physics of the National Academy of Sciences of Ukraine, 46 Nauky Ave., 03039 Kyiv, Ukraine}
\author{Marcin Jarzyna}\thanks{e-mail: m.jarzyna@cent.uw.edu.pl}
\affiliation {Centre for Quantum Optical Technologies, Centre of New Technologies, University of Warsaw, Banacha 2c, 02-097 Warszawa, Poland 
}

%markboth{Journal of Lightwave Technology}%
%{Uzunova \MakeLowercase{\textit{et al.}}: Photon Efficiency of High-dimensional Quantum Key Distribution}

\begin{abstract}
We investigate entanglement-based quantum key distribution protocols, with particular emphasis on their efficiency under realistic conditions of satellite quantum communications, where performance is limited by the low power of a received signal and background radiation. We focus on scenarios where each photon pair is used to encode multiple qubits in order to optimally utilize the weak signal. By optimizing over the source intensity and the number of encoded qubits we study the theoretical information limit for the QKD efficiency. We show that the optimal efficiency is attained for finite entangled photons pair production probability which is in contrast to conventional communication efficiency maximized in the limit of vanishing signal strength. The multiqubit encoding can enhance the secret key rate by up to an order of magnitude compared to single-qubit schemes.
\end{abstract}

\maketitle

\section{Introduction}

Cryptography lies at the heart of modern communication, constantly evolving to meet the requirements of privacy of storage, processing and transmission of sensitive data. The scope of applications includes sensitive sectors such as healthcare, critical infrastructure, telecommunication, financial services, and government agencies \cite{Aquina2025}. In general cryptographic methods ensure the confidentiality of information transmitted over insecure communication channels by encoding the sensitive data through a sequence of bits, known as a cryptographic key, which allows access only to authorized parties. Without knowledge of the secret key, the data cannot be decrypted. A central challenge in cryptography is the secure distribution of this secret key between two legitimate users—conventionally referred to as Alice and Bob—while preventing any information leakage to a potential eavesdropper, Eve.

Quantum key distribution (QKD) promises to revolutionize cryptography through incorporation of laws of quantum mechanics in the security paradigm. The security of QKD originates in the fundamental no-cloning theorem which states that it is impossible to perfectly copy an unknown quantum state \cite{Wootters1982NoCloning}. Thus, a potential eavesdropper, Eve, cannot gain enough information about a transmission between Alice and Bob without altering it in a random and uncontrollable way likely to be detected. Crucially, this means that security of QKD is based not on computational difficulty of certain tasks, typical for classical methods \cite{diffie1976new, RSA}, such as large number factorization which can be threatened by e.g. quantum computers \cite{Shor1994, arute2019quantum}, but only on the underlying laws of physics. One can therefore argue about \textit{unconditional} security of QKD, since in principle it cannot be broken by any computational assumptions etc. \cite{BBM92, ShorPreskill2000}. Note, however, that in practice this is limited by ones assumptions about the utilized quantum channel and possible unknown side-channels that can be exploited by untrusted parties.

One of the main challenges of QKD is to extend the quantum communication range to large distances and ultimately up to a global scale. Despite recent progress \cite{korzh2015provably, boaron2018secure, pistoia2022paving}, current fiber optical demonstrations are usually limited to tens or few hundreds of kilometers \cite{neumann2022continuous, honjo2008long} and require separated fiber network, which make them exceptionally expensive \cite{Aquina2025}. One solution to this problem is to use satellite and free-space QKD \cite{liao2017satellite, bedington2017progress} which relies on the much higher optical transmission of the atmospheric channel than fibers. Successful implementation of QKD for free-space optical links allowed to distribute a secure key over hundreds and even thousand km range \cite{SchmittManderbach2007,Ursin2007,YinScience2017}, including entanglement-based key distribution \cite{YinNature2020}.

Entanglement based QKD allows to extend to range of key distribution \cite{E91} and to let go the assumption of trusted nodes, thus further increasing the security. However, entangled photon sources are typically characterized by low brightness \cite{kwiat1995high_intensity_entangled, OptCom2025Bright, Park2024Ultrabright}, which, combined with significant losses due to atmospheric propagation, lead to a very low received signal strength. In order to obtain optimal performance of such QKD link it is therefore crucial to extract as much key as possible per single received photon. This is a similar situation to classical communication operating in photon starved regime \cite{hemmati2006deep} where the valid figure of merit is photon information efficiency (PIE). Motivated by the latter one should quantify the performance of a QKD link by analogous quantity rather than just the key rate.

The majority of current photonic free-space quantum communication systems, including recent quantum satellite experiments \cite{Li2025}, use two-dimensional encoding e.g. in polarization, where each photon can carry at most a single bit of quantum information. However, spatio-temporal or other degrees of freedom of a photon can be used for high-dimensional encoding, in essence realizing multiple qubits transmission by a single photon \cite{ZwolinskiECOC2020,BanaszekJachuraICSOS2017}. Employment of  high-dimensional quantum states could increase QKD key rate and improve robustness with respect to noise and eavesdropping \cite{Sulimany2022, Yu2025,sit:17,ChenHabifNPH2012}.

In this paper, we consider the implementation of high-dimensional encoding for known entanglement-based QKD protocols and investigate the theoretical limits of secret key photon efficiency. The paper is organized as follows.  Sec.~\ref{sec:security} gives introduction to QKD. The multiqubit encoding and reception scheme is described in Sec~\ref{sec:multi} and the corresponding QKD link model in Sec~\ref{sec:link_model}.  Next, Sec. \ref{sec:opt} contains results on optimizing QKD photon efficiency. Finally, Sec. \ref{sec:conc} concludes the paper.

 \section{Quantum Key Distribution\label{sec:security}}
The goal of QKD is to establish a secret key between two parties, Alice and Bob.  In entanglement based approach, both parties receive one of two photons from an entangled pair that was emitted by a separate light source. They then each perform measurements independently in preselected random bases. By means of classical communication they then reject events that are obviously uncorrelated, e.g. transmitter and receiver bases were different, and perform error correction and privacy amplification on the remaining bit sequences. The presence of any kind of noise, errors or other unwanted effects is prescribed to an action of a possible eavesdropper, Eve.
In this paper we consider $s=4$ four- and $s=6$ six-state BBM92 protocols \cite{BBM92}, \cite{BB84six}, as well as entanglement-based versions of the four- and six-state SARG04 protocols \cite{Sarg04},\cite{SARG04six}, which are described in detail in App.~\ref{app:protocols}.

The theoretical asymptotic secret key rate $K$ of any QKD protocol is given by \cite{Devetak}
\begin{equation}\label{eq:key}
    K=I(A,B)-\chi(B,E),
\end{equation}
where $I(A,B)$ is the mutual information between Alice and Bob, $\chi(B,E)$ is the Holevo information between Bob and Eve, quantifying the maximum possible information that Eve can posses on Bob's results allowed by the laws of quantum mechanics, and we assumed classical data in the postprocessing to be send from Bob.
We assume measurements performed in eigenbases of Pauli $X, Y$ and $Z$ operators, with respective quantum bit error rates (QBER) denoted by $e_X, e_Y, e_Z$. 

The value of QBER in different bases can be associated with projections of Alice's and Bob's shared bipartite entangled state $\rho_{AB}$ onto different Bell states \cite{asymmetric} through set of equations
\begin{align}\label{eq:ppp}
p_X=&\bra{\Psi^+}\rho_{AB}\ket{\Psi^+}=\frac{1}{2}(e_Z-e_X+e_Y),\\
p_Y=&\bra{\Psi^-}\rho_{AB}\ket{\Psi^-}=\frac{1}{2}(e_Z+e_X-e_Y),\notag\\
p_Z=&\bra{\Phi^-}\rho_{AB}\ket{\Phi^-}=\frac{1}{2}(e_X+e_Y-e_Z),\notag\\
p_I=&\bra{\Phi^+}\rho_{AB}\ket{\Phi^+}=1-p_X-p_Y-p_Z,\notag%\frac{1}{2}(e_X+e_Y+e_Z),\notag
\end{align}
where 
the Bell states $\ket{\Phi^{\pm}}$ and $\ket{\Psi^{\pm}}$ are given by
\begin{align}
\ket{\Phi^\pm}=\frac{1}{\sqrt{2}}\left(\ket{0}_A\ket{0}_B\pm\ket{1}_A\ket{1}_B\right), \\ \ket{\Psi^\pm}=\frac{1}{\sqrt{2}}\left(\ket{0}_A\ket{1}_B\pm\ket{1}_A\ket{0}_B\right)\notag,
\end{align}
and we assumed the source generates photons in the $\ket{\Phi^+}$ state. 
The secret key fraction in Eq.~\eqref{eq:key} can be then expressed as \cite{ScaraniRMP2009}
\begin{align}\label{eq:key_p}
K=1+\sum_{i\in\{I,X,Y,Z\}}p_i\log_2p_i,%\;\;i=I,X,Y,Z,
\end{align} 
only one basis to generate the key.
One can relate the above probabilities to various types of errors in an entanglement distillation task 
\cite{ShorPreskill2000,LoChau1999}. Specifically, $p_I$ denotes probability of no error, $p_X$ represents probability of a bit flip, $p_Z$ the probability of a phase flip and $p_Y$ is the probability of simultaneous bit and phase flip. {Note, that in some protocols, such as four-state BBM92 one does not perform measurements in all three $X, Y, Z$ bases. In such case one needs to minimize the key formula in Eq.~\eqref{eq:key_p} over QBER corresponding to the unobserved basis.} The total probability of a bit error is then given by $e_{\textrm{bit}}=p_X+p_Y$ and for phase error one obtains $e_{\textrm{ph}}=p_Z+p_Y$. 

For a protocol specified by given values of bit and phase errors probabilities,
the key rate in Eq.~\eqref{eq:key_p} is given by
\begin{equation}\label{eq:rr}
K=1-h(e_{\textrm{bit}})-h(e_{\textrm{ph}}|e_{\textrm{bit}}),
\end{equation}
{where $h(x)=-x \log_2 x (1-x)\log_2(1-x)$ is the binary entropy function and } the second term in Eq.~\eqref{eq:rr}, $h(e_{\textrm{ph}}|e_{\textrm{bit}})$, is the Shannon entropy of the phase error conditional on the bit error pattern.

In practice Alice and Bob can use their measurement bases with unequal probabilities. For example, assuming the fraction of measurements in the $Z$ basis is denoted by $q$ for the four-state BBM92 protocol using $X$ and $Z$ bases the average QBER is $qe_Z + (1-q)e_X$. However, in such instance, instead of standard approach which uses both bases for key one may use $Z$ basis to produce the key while measurements in the $X$ basis serve just for QBER estimation. This is particularly beneficial when the effective quantum channel between Alice and Bob affects the basses in an asymmetric way. For the aformentioned BBM92 protocol one obtains the key rate
 \begin{equation}\label{bbm92}
     K^{\textrm{BBM92}}_{s=4}=q\left[1-h(e_{Z})-h(e_{X})\right].
 \end{equation}
Note that $q$ only affects the statistical accuracy of the QBERs estimates accounting the effects of the final key.

\begin{figure}[t]
  \includegraphics[width=1\linewidth]{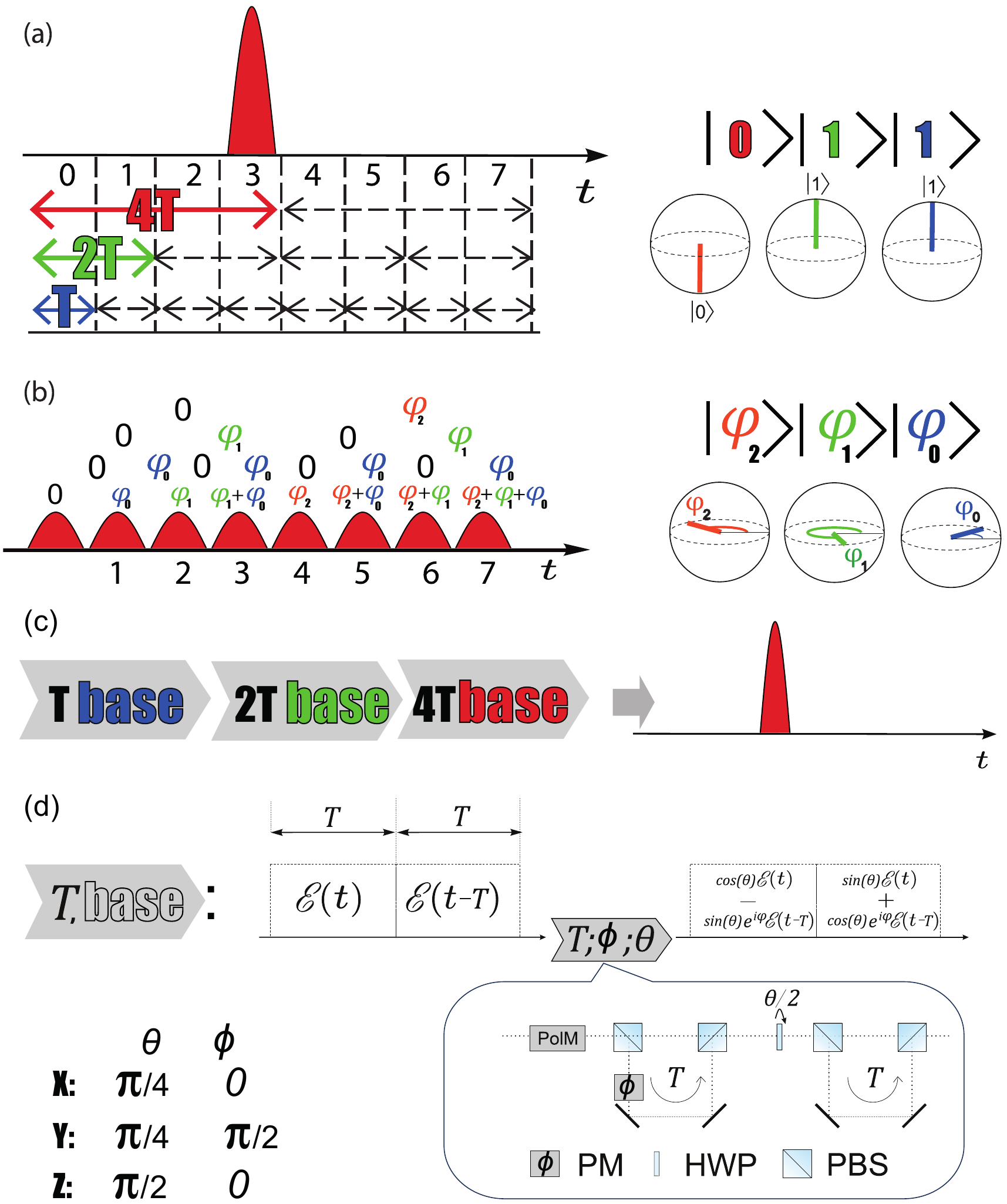}
  \caption{Encoding of multiple logical qubits into a single photon prepared as a superposition of temporal slots. Example is given for three qubits, $m=3$. (a) The slot number written in the binary representation specifies the basis states of individual qubits. The presence of a photon in  temporal slot with $n=3$ corresponds to a product of qubit basis states $\ket{0}\ket{1}\ket{1}$ in computational basis.  (b) The classical sequence of three bits can be equivalently encoded in 3 equatorial qubits states $\ket{\phi_{2}}\ket{\phi_1}\ket{\phi_0}$ by a light pulse in the time frame with the corresponding relative phase in each of the $M$ slots. Transformation
introduced by a cascade of interferometric stages (c). With a suitable choice of beam splitter parameters, orthogonal pairs of
states of logical qubits are bijectively mapped onto the temporal position of the photon at the output of the cascade. (d)  An interferometric stage acting as a beam splitter combining the optical field in two adjacent time intervals of
duration $T$. The parameters $\theta$ and $\Phi$ determine the field transformation realized by the beam splitter.  PM, phase modulator; HWP, half-wave plate; PBS, polarizing beam splitter.}
\label{fig1}
\end{figure}

\section{Multiqubit encoding}
\label{sec:multi}

Satellite-based optical communication is associated with inevitable significant losses. This, combined with the fact that typical entanglement sources are characterized by a small pair production probability of $p_{\textrm{pair}}\ll 1$ \cite{kwiat1995high_intensity_entangled, OptCom2025Bright, Park2024Ultrabright} leads to very weak signal strength at the receiver. As already mentioned in the introduction, in such a regime, one wants to maximize the amount of secret key bits that can be extracted per single received photon since the latter is a precious resource. One way to do this is to encode multiple qubits in a single photon since then each detection event could in the ideal case yield multiple bits of secret key. Below we will describe one such method proposed in \cite{Jachura2021}.

\subsection*{Signal encoding}

Consider a single photon occupying one of $M=2^m$ temporal slots together composing a frame of duration $\Delta t$. One may write the index of the photon-carrying slot $k$ in binary representation as $k=k_{m-1}\dots k_1k_0$ where $k_i=0,1$. In such a representation the state of a photon occupying the $k$-th slot naturally corresponds to a combination of computational basis states of $m$ qubits $\ket{k}\sim\ket{k_{m-1}}\dots\ket{k_1}\ket{k_0}$ ,see Fig.~\ref{fig1}(a). This encoding is similar to  the pulse position modulation format, which is standard in classical communication.

In order to encode an arbitrary state of $m$ qubits, however, one needs to be able to represent in a similar manner also all possible superpositions of basis states. This can be easily done by taking superpositions of states representing a photon in a slot corresponding to each individual term in $m$-qubits superposition expressed in computational basis with appropriate phase factors. For example if one wants to encode an equal superposition of $m$ qubits states $\ket{\psi}=\otimes_{j=1}^m \left(\ket{0}+e^{i\varphi_j}\ket{1}\right)/\sqrt{2}$ the result would be a photon distributed uniformly over all slots with each slot in the superposition carrying a phase equal to sum of phases of individual qubits for which 1 appears in the binary representation of the slot index
\begin{equation}
    \ket{\psi}\sim \sum_{k=0}^{2^m}e^{i\sum_{j=0}^{m-1}\varphi_jk_j}\ket{k},
\end{equation}
where $k=k_{m-1}\dots k_0$ in binary representation for every $k$. The encoding is presented schematically in Fig.~\ref{fig1}(b). The procedure can be naturally extended to the entangled case by considering an entangled photon pair and performing encoding on each photon separately. This results in a two photon state with one-to-one correlations between individual temporal time slots.

\subsection*{Receiver scheme}

In order to decode the multiqubit signal, either in a standard or an entangled case, the receiver maps the photons incoming in the whole $\Delta t$ time frame into PPM format. The transformation can be accomplished by a  cascade of interferometric stages combined with active polarization switching,  similarly to the receiver architecture proposed for photon-efficient classical communication \cite{ZwolinskiECOC2020,BanaszekJachuraICSOS2017,GuhaPRL2011}. 
For $m$ qubits the setup is composed of $m$ stages, shown schematically in Fig.~\ref{fig1}(c,d). Each module combines the optical field from pairs of adjacent time intervals of duration $2^{i}T$, where $i=0,..,m-1$ enumerates the stages (and qubits) and $T$ is the duration of a single slot, $T=\Delta t/M$. This is accomplished by first actively switching the polarization of light in the first interval from a pair by a polarization modulator. Next, the signal is divided on a polarizing beam splitter and one of the beams is delayed in order to occupy the same time interval as the second beam. The two beams are then recombined on another polarizing beam splitter which is followed by a half-wave plate (HWP) rotating the polarization by $\theta$, which determines the choice of the measurement basis for the particular qubit. Finally, light passes through another interferometer to redistribute the possible occupied slots on the whole initial interval. Intuitively, each module acts as a beam splitter in the temporal domain, reducing the length of the sequence by a factor of two and doubling the intensity of light in individual slots.  At the output the optical energy is concentrated in a single temporal slot, caring information about quantum states of all $m$ qubits. The overall transformation allows to perform measurements in a basis $\ket{\phi_{+}}=\sin\theta\ket{0} + e^{i\phi}\cos\theta\ket{1},\,\ket{\phi_{-}}=\cos\theta\ket{0} - e^{i\phi}\sin\theta\ket{1}$.  To measure qubit in $X$ basis one needs to set $\theta=\pi/4$ and $\Phi=0$, for $Y$ basis $\theta=\pi/4$ and $\Phi=\pi/2$, and  $\theta=\pi/2$, $\Phi=0$ for $Z$ basis. Importantly, the measurement basis can be selected independently for each of the logical qubits by an appropriate choice of the phase and the splitting ratio of the corresponding stage. The final detection needs to identify the position of the light pulse within a sequence of empty time slots. This can be easily done by direct detection from the timing of photocount events. Instances in which no detection is recorded over the entire $\Delta t$ time frame are interpreter as erasures. Importantly, note that in the entangled scenario if Alice and Bob chose the same basis for the corresponding qubit in their respective interferometric stages, their results will be correlated, allowing them to extract secret key.

\section{Multiqubit Satellite Link Model}
\label{sec:link_model}

Let us assume the source produces entangled photon pairs at a rate $R_{\textrm{source}}=p_{\textrm{pair}} /\Delta t$, where $p_{\textrm{pair}}$ is the probability of generating a photon pair within the time frame $\Delta t$, and that one encodes $m$ qubits into them. In a realistic setting, each photon experiences non-ideal transmission, specified by $\eta_A,\,\eta_B$ for the Alice's and Bob's paths, respectively, and background radiation with respective strengths $n_A,\,n_B$. The receiver may also be a source of additional noise which may be conveniently characterized by introducing the disturbance $D$ which is a fraction of incorrect outcomes introduced just by the particular measurement apparatus, i.e. assuming there is no other source of errors. 
In a similar manner one may introduce also the fraction of conclusive events $E$ which denotes how many detected photon pairs contribute in the actual key. The probability of {a conclusive} event $E$ is given by $E^{\textrm{BBM92}}=1$ for BBM92 and $E^{\textrm{SARG04}}=D+\frac{1}{2}$ for SARG04 protocol.

\begin{figure*}[t]
  \includegraphics[width=2\columnwidth]{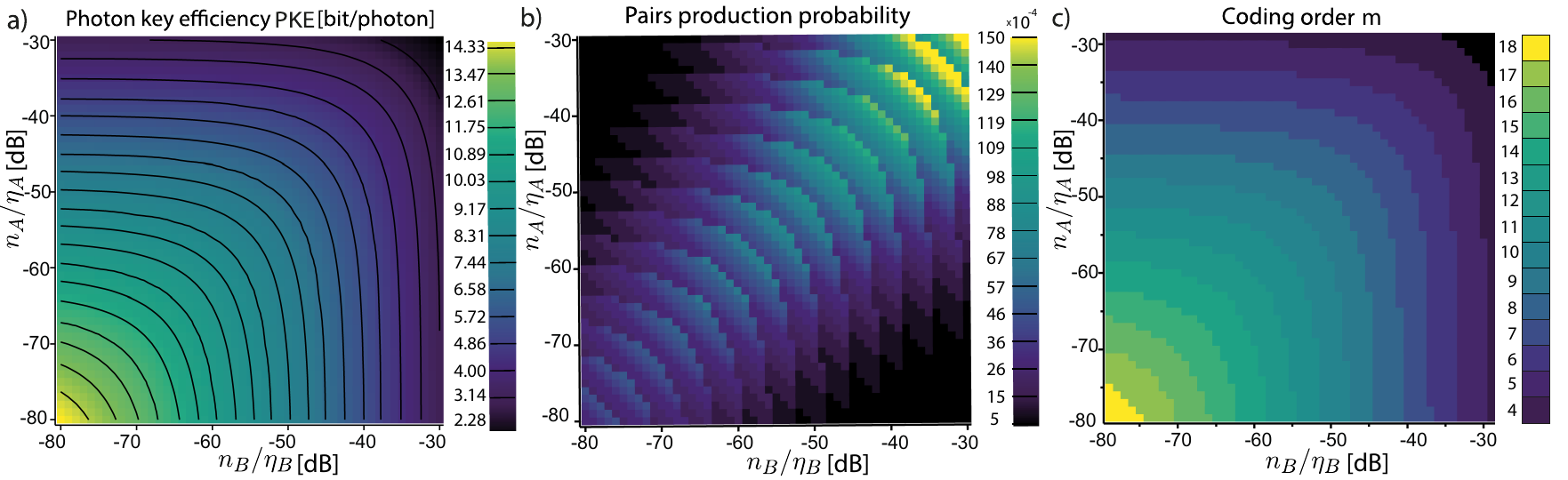}
  \caption{Optimal PKE (a), the corresponding the pair-production probability (b) and (c) coding order, as functions of the decoherence parameters in Alice’s and Bob’s channels $n_A/\eta_A$ and $n_B/\eta_B$ for the asymmetric four-state BBM92 protocol and visibility parameter is $V=0.98$.}
  \label{fig:contour}
\end{figure*}

The total key rate $R$ is given by the product of the key rate per single event per single qubit in Eq.~\eqref{eq:rr}, the events rate $R_{\textrm{event}}$ registered by the receiver and the number of encoded qubits $m$ i.e. $R_{\textrm{key}}=m R_{\textrm{event}}K$. The latter, assuming small pair production probability $p_{\textrm{pair}}\ll 1$ and weak noise $n_A, n_B\ll 1$, in the leading order is composed of several different contributions listed below. 

    \paragraph{Coincidences between pairs of signal photons} The rate of two photons passing through imperfect channels to Alice and Bob is given by $p_{\textrm{pair}}\eta_A\eta_B/\Delta t=R_{\textrm{source}}\eta_A\eta_B$. In such instances  the probability of obtaining a conclusive result of the protocol regardless of its correctness is {equal to} $E$ while the probability of error is $D$.
     \paragraph{Coincidences between signal and background photons} If one of the registered photons originates from the background radiation the corresponding rate is given by $2^mn_AR_{\textrm{source}}\eta_B$ if the noisy photon was registered by Alice and $2^mn_BR_{\textrm{source}}\eta_A$ if it was registered by Bob. Since the signal and noisy photons are uncorrelated the probability of incorrect result is equal to $1/2$.
      \paragraph{Coincidences between pairs of background photons} In case both registered photons came from the background radiation the corresponding event rate is given by $2^{2m} \Delta t=2^{2m}R_{\textrm{source}}/p_{\textrm{pair}}$ and, since they are uncorrelated, the incorrect result probability is $1/2$.
    \paragraph{Double pair events} It may rarely happen that two signal pairs are generated in the same time interval $\Delta t$. The rate corresponding to such events is equal to $2\eta_A\eta_BR_{\textrm{source}} p_{\textrm{pair}}$. If the registered photons belong to the same pair, which happens in half of the cases, the probability of a conclusive result of the protocol is equal to $E$ and the probability of error is $D$.  Otherwise, if the registered pair is uncorrelated, the probability of a conclusive result is $1$ and the probabilities of correct and incorrect results are equal to $1/2$. The total probability of a conclusive result is therefore given by $(E+1)/2$ and for the probability of error $(D+1/2)/2$.

\begin{figure*}[t]
  \includegraphics[width=2\columnwidth]{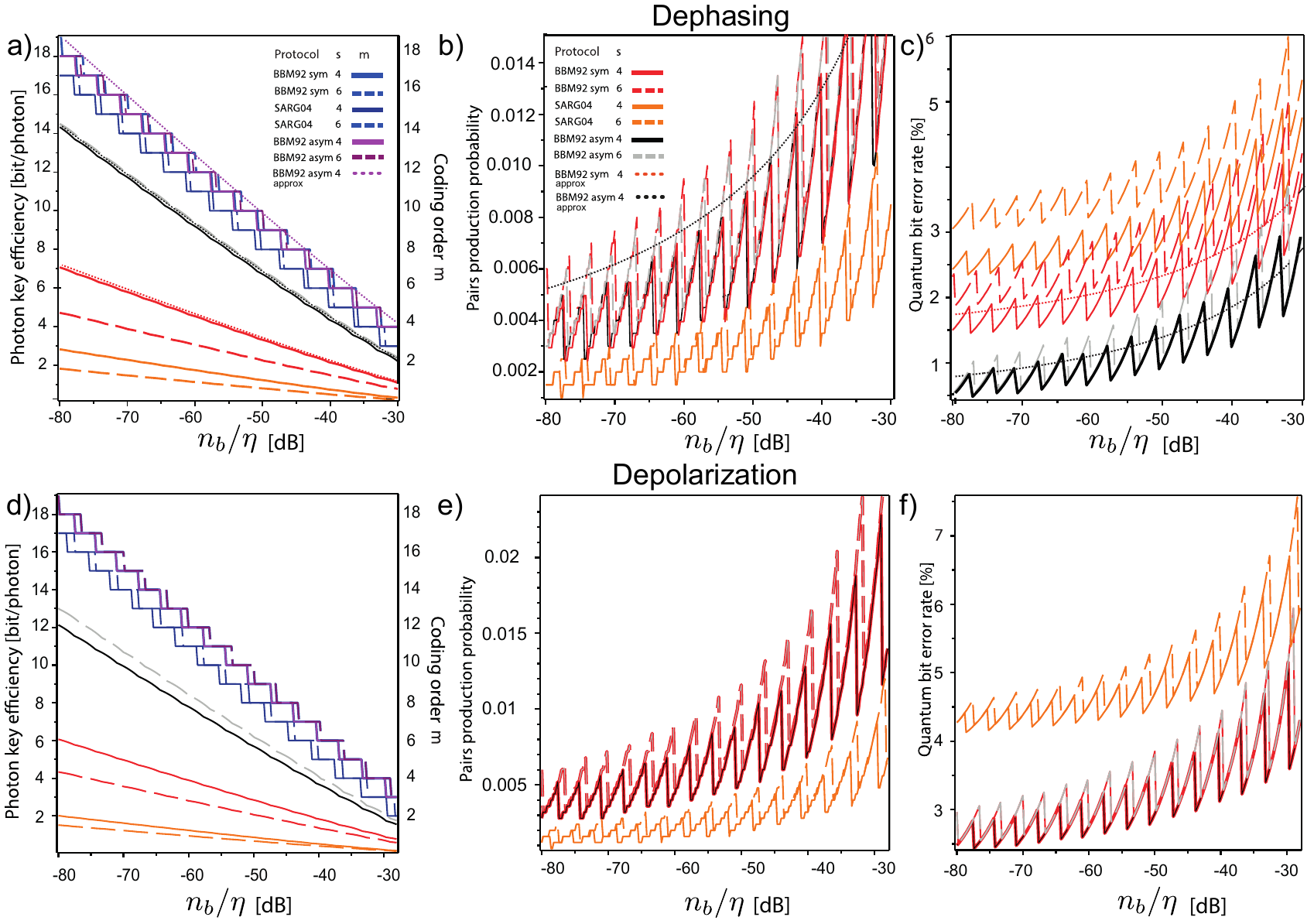}
  \caption{Optimal photon key efficiency (PKE) (a, d), the corresponding quantum bit error rate (QBER) (b, e), and the optimal pair-production probability (c, f) for different protocols. Panels (a–c) correspond to the dephasing channel model, while panels (d–f) correspond to the depolarizing channel model. The visibility parameter used for the calculation is $V=0.98$. Thin black dashed lines show approximate analytical results.}
\label{fig3}
\end{figure*}

The combined rate of detection events is given by 
\begin{align}
R_{\textrm{event}}^{}=R_{\textrm{source}}\eta_A\eta_B\left[E+M\left(\frac{n_A}{\eta_A}+\frac{n_B}{\eta_B}\right)+\right. \\ +\left.\frac{M^2}{p_{\textrm{pair}}}\frac{n_A n_B}{\eta_A\eta_B}+\left(E+1\right)p_{\textrm{pair}}\right],\notag
\end{align}
and the error events
\begin{align}
R_{\textrm{error}}^{}=R_{\textrm{source}}\eta_A\eta_B\left[D+\frac{1}{2}M\left(\frac{n_A}{\eta_A}+\frac{n_B}{\eta_B}\right)+\right. \\ +\left.\frac{1}{2}\frac{M^2}{p_{\textrm{pair}}}\frac{n_A n_B}{\eta_A\eta_B}+\left(D+\frac{1}{2}\right)p_{\textrm{pair}}\right].\notag
\end{align}
The QBER is given by the ratio $R_{\textrm{error}}/R_{\textrm{event}}$. The exact value of error probability $D$ depends on the decoherence experienced by the receiver. We consider two types of decoherence: depolarization and dephasing. The former describes uniform loss of coherence across all qubit bases and typically models diverse noise sources like pointing jitter (mechanical vibrations and tracking errors) and atmospheric turbulence.  The latter is often used as a mathematical model of imperfect interferometer with non-unit visibility, quantified by visibilities $V_A$ and $V_B$ for Alice's and Bob's receivers, which is a crucial detrimental effect for multiqubit encoding and reception. Both decoherence processes are described in detail in App.~\ref{app:noise}.

In practical applications the pair production probability is small and, as discussed above, one should maximize the amount of secret key bits that can be extracted from a single received photon pair. The correct figure of merit is then photon key efficiency (PKE)
\begin{equation}\label{eq:PKE}
\textrm{PKE}=\frac{R_{\textrm{key}}}{\eta_A\eta_B R_{\textrm{source}}}=\frac{mR_{\textrm{event}}K}{\eta_A\eta_B R_{\textrm{key}}},
\end{equation}
i.e. the ratio of the total key rate to the received signal rate. Note that PKE depends not only on the quantum channel and measurement but crucially also on the protocol.

\section{Optimal photon efficiency}
\label{sec:opt}

In order to obtain maximal PKE one can optimize it with respect to the number of encoded qubits in a single photon $m$ and the source intensity, quantified by the pair production probability $p_{\textrm{pair}}$. To characterize decoherence introduced by the channel, one may use the noise parameter given by the noise to transmission ratio for each party $n_A/\eta_A$ and $\eta_B/n_B$. It is seen in Fig.~\ref{fig:contour}(a) that PKE is symmetrical with respect to Alice's and Bob's noise parameters. The optimal pair production probability depicted in Fig.~\ref{fig:contour}(b) decreases with decreasing noise parameter while the optimal coding order in Fig.~\ref{fig:contour}(c) increases.

\begin{figure}[t]
   \centerline{\includegraphics[width=\linewidth]{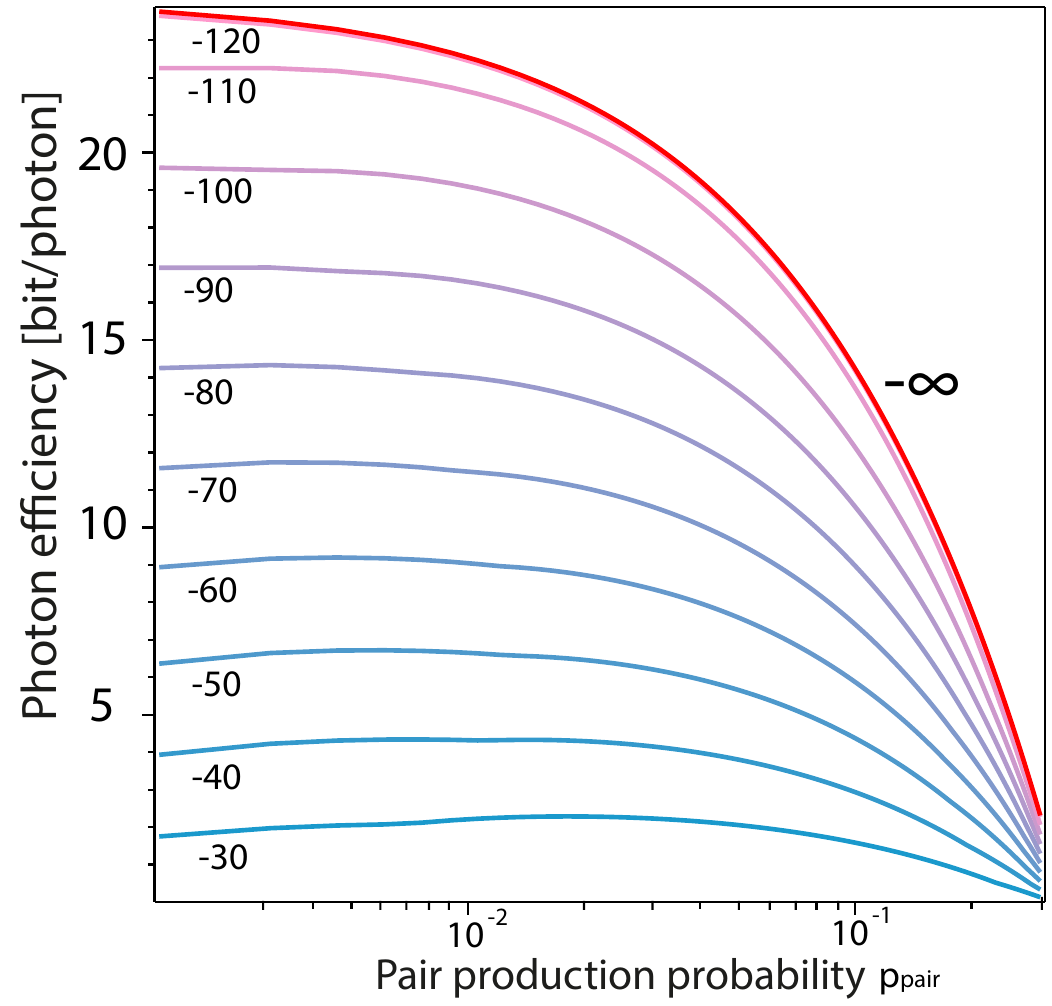}}
      \caption{\label{fig:PKE_pair} Photon key efficiency for asymmetric four-state BBM92 protocol with $q=1$ as a function of pair production probability assuming dephasing channel with visibility $V=0.98$. Numbers indicate the values of the noise parameter $n_b/\eta$.}
\end{figure}

Optimal PKE and QBER for different entangled-based protocols and decoherence models assuming the same channels on Alice's and Bob's side, $\eta_A=\eta_B=\eta$, $n_A=n_B=n_b$ and $V_A=V_B=V$ is presented in Fig.~\ref{fig3}. It is seen in Figs.~\ref{fig3}(a), (d) that  the best possible photon efficiency can be obtained using BBM92 protocol. The efficiency is further increased if Alice and Bob measure their photons with a biased choice of basis  i.e. $q\approx 1$, meaning they predominantly use the $Z$ basis, with only small fraction of events corresponding to $X$ basis used to determine QBER. It is seen also that the optimal number of qubits encoded in a single photon scales logarithmically with the noise parameter $n_b/\eta$. Figs.~\ref{fig3}(b,e) illustrate the corresponding values of QBER for different protocols as a function of the noise parameter $n_b/\eta$ and Figs.~\ref{fig3}(c,f) depict the corresponding optimal pair production probability.  It is seen that  the optimal pair production probability depends on the protocol and is about two times smaller for SARG04 than for BBM92. It decreases slowly with the decoherence parameter, eventually reaching $0$ if there is no external noise $n_b=0$. The PKE as a function of pair production probability for different levels of noise parameter is seen in Fig.~\ref{fig:PKE_pair}. It is seen that optimal photon efficiency increases as noise is attenuated, eventually approaching the noiseless case. Note that in the noisy case the PKE is in general a quite flat function and its maximum is not well pronounced. The noiseless case exhibits no maximum, i.e. PKE grows indefinitely as pair production probability decreases, however, the growth is very slow.

In case of a weak noise $n_b/\eta \ll 1$, one can find an approximate analytical behavior for PKE an the optimal pair production probability, given by
\begin{equation}
\textrm{PKE}=q\left[1-2h\left(D+\frac{\bar{p}_{\textrm{pair}}}{2}\right)\right]\frac{[W(\Xi)-1]^2}{W(\Xi)\ln{2}},
\end{equation}
where $W(x)$ is the Lambert W function and the parameter $\Xi$ is given by
\begin{equation}\label{eq:Xi}
\Xi=\frac{e\eta}{2n_b}\frac{2h\left(D\right)-1}{\log_2\left[2D\left(1-{D}\right)\right]},
\end{equation}
where $D=(D_X+D_Z)/2$ is the average disturbance. The optimal number of qubits encoded in a single photon $m^{*}$ treated for simplicity as a continuous parameter, is equal to
\begin{align}\label{eq:m}
m^{*}=\frac{W(\Xi)-1}{\ln{2}},
\end{align}
and the optimal probability of generating a pair of photons
\begin{equation}\label{eq:p}
p^{*}_{\textrm{pair}}=\frac{1}{2}\left(\frac{e\eta W(\Xi)}{2n_b\Xi}-1\right)^{-1}.
\end{equation}
The details of calculations are presented in Appendix~~\ref{app:PKE}. It is seen in Fig.~\ref{fig3} that the approximation agrees with numerically obtained values of PKE, QBER and $p_{\textrm{pair}}$ in the regime of weak noise.

In the regime of low noise parameter $n_b/\eta\ll 1$ the optimal performance of the protocol is achieved at low source intensities. This is similar to classical communication, where the optimal photon information efficiency for the additive noise channel is attained in the limit of vanishing signal strength \cite{Verdu1990, Jarzyna2017, Ding2017}. The same behavior can be observed for QKD, when there is no background noise $n_b=0$ since then $\Xi\to\infty$ in Eq.~\eqref{eq:Xi} and $p^{*}_{\textrm{pair}}\to 0$ in Eq.~\eqref{eq:p}, giving, similarly as in the classical scenario \cite{GuhaPRL2011} $\textrm{PKE}\to\infty$. However, in contrast to classical case, when the decoherence parameter is nonzero $n_b/\eta>0$, the optimal pair production rate remains small but finite. This is because the qualitatively different role of the background noise in QKD, which is much more fragile. In particular, if the noise is stronger than the signal it is impossible to generate the secret key, which imposes a lower bound on the allowed pair production rate $p_{\textrm{pair}}>2^mn_b/2\eta$. The noise also imposes a limitation on the number of qubits that can be encoded in a single photon. Specifically, the number of temporal slots must satisfy $M<\eta/2n_b$, and the maximal number of encoded qubits is given by $\log_2\eta/2n_b$. For example, under daylight conditions, characterized by the noise  $n_b=10^{-4}$ and transmittance $\eta=10^{-3}$ values, only two qubits can be encoded.

\section{Conclusions \label{sec:conc}}
Satellite QKD provides a way to establish a secure key
for cryptographic purposes without deploying costly terrestrial network infrastructure. We have theoretically investigated a high-dimensional quantum key distribution scheme based on entangled photon pairs where the information is encoded in the  arrival time of photons and the relative phases between them. We obtained that high-dimensional encoding significantly enhances the secret key rate per photon and improves robustness against noise possibly originating from eavesdropping attacks. For unrestricted signal bandwidth, the attainable photon efficiency is effectively limited by the background noise acquired by the propagating optical signal. In contrast to conventional communication, where the optimal photon information efficiency is attained for received signal power converging to zero, in QKD the optimal signal strength exhibits a clear non-vanishing maximum. One can also view optimal value of photon key efficiency as quantification of fundamental energy efficiency of a given QKD protocol, however, note that it only accounts for the energy cost of signal generation and does not take into account energy necessary for measurement, postprocessing etc. \cite{Yehia2025}.

\section*{Acknowledgment}

The authors would like to thank Konrad Banaszek for insightful discussions. 
This work was supported by the European Union’s Horizon Europe research and innovation programme under the project ‘Quantum Security Networks Partnership’ (QSNP, Grant Agreement No. 101114043).
\appendix
%\appendices

\section{QKD protocols}
\label{app:protocols}

Below we will briefly describe BBM92 and entanglement based version of SARG04 protocols.

\subsection{BBM92 protocol}
We consider the four- and six-state versions of the BBM92 protocol, utilizing $N=2$ and $N=3$ measurement bases respectively. Both versions are composed of the following steps:
\begin{itemize}

\item{For each time frame, Alice and Bob perform a measurement randomly in one of $N$ bases.}

\item{For each detected photon Bob} publicly announces the basis he used in his measurement.

\item{Alice and Bob compare their choices of bases and discard all the measurement results where they used a different basis. The ratio of discarded bits is $(N-1)/N$. The remaining bit string forms the sifted key. }

\end{itemize}

\subsection{SARG04 protocol}

The four-state SARG04 protocol utilizes two bases \mj{typically Z, $\{\ket{0},\ket{1}\}$, and X, $\{\ket{+},\ket{-}\}$, where $\ket{\pm}=(\ket{0}\pm\ket{1})/\sqrt{2}$.} 
Steps of the SARG04 are as follows:

\begin{itemize}
 \item{The source produces a sequence of entangled photon pairs described by a Bell state $\ket{\Phi^+}$ and sends one photon from the pair to Alice and Bob.}

\item{Alice and Bob each perform a measurement choosing randomly one of the two bases.}

\item{
For each detected photon pair, Alice publicly announces one of the two
classical sets containing her measurement result. The classical sets are represented by pairs of Z and X states $\{0,-\}$, $\{-,1\}$, $\{1,+\}$, $\{+,0\}$.}

{
\item Bob compares his measurement outcome to the classical set sent by Alice. If it is orthogonal to one of the states mentioned by Alice, he concludes that the other symbol in the classical set is the correct result. Otherwise, if Bob's outcome  is explicitly mentioned in the classical set sent by Alice, the result is inconclusive. The probability of a conclusive event is given by $E^{\textrm{SARG04}}=D+1/2$.}

\end{itemize}
The six-state SARG04 protocol utilizes  three bases $X$, $Y$ and $Z$ and twelve corresponding classical sets \cite{SARG04six}.

\section{Decoherence effects in the receiver}
\label{app:noise}

An arbitrary quantum channel describing decoherence process is characterized by two sets of Kraus operators, $\{K_A^i\}$ for Alice's and $\{K_B^i\}$ for Bob's channels respectively. The action of the two-sided channel on an arbitrary input state $\rho_{AB}$ is given by
\begin{equation}
\varepsilon{(\rho_{AB})}=\sum_{i,j}K_i^AK_j^B\rho_{AB} K_i^{A\dagger}K_j^{B\dagger}.
\end{equation}
It is convenient to introduce two quantities to characterize the quality of the output quantum state between Alice and Bob, $F_\phi$ and disturbance $D_\phi$  defined  as 
\begin{equation}\label{D}
D_\phi= \frac{\bra{\phi}_A\bra{ \phi_\bot}_B[\varepsilon{(\rho)}]\ket{\phi}_A\ket{ \phi_\bot}_B}{\bra{\phi}_A \Tr_B[\varepsilon{(\rho)}]\ket{\phi}_A},\, F_\phi=1-D_\phi.
\end{equation}
for an orthonormal basis $\{\ket{\phi},\ket{\phi_\bot}\}$. If both Alice and Bob perform a measurement in this basis, they will both  get $\ket{\phi}$ with probability $F_\phi$ and with probability $D_\phi$ Alice will observe $\ket{\psi}$ and Bob $\ket{\psi_\bot}$.

\subsubsection*{Dephasing noise}
The Kraus operators for dephasing channel are {given by}
\begin{equation}\label{kdeph}
K^c_0=\sqrt{\frac{1+V_c}{2}}I,\;\;K^c_1=\sqrt{\frac{1-V_c}{2}}\sigma_Z
\end{equation}
where the dephasing parameters $V_c$, $c=A,B$ for Alice and Bob channels respectively can be interpreted as visibilities of their interferometers and $\sigma_Z$ is the Pauli Z matrix. 

Crucially, the dephasing channel is asymmetric and can be interpreted as reduction of the Bloch ball along $X$ and $Y$ directions, transforming it into an ellipsoid \cite{breuer2002theory}. Note that this means that the effect of dephasing may differ depending on the input state.
Specifically, if the input is one of the $Z$ basis states, $\{\ket{0},\ket{1}\}$, the channel acts as identity, resulting in zero error probability, $F_Z=1,\;D_Z=0$. On the other hand, if the input is one of $X$ or $Y$-basis states, which are influenced the most by dephasing, the probabilities are similar given by
\begin{equation}
 F_X= F_Y=\frac{1+V_AV_B}{2};\;\; D_X= D_Y=\frac{1-V_AV_B}{2}.
\end{equation}

\subsubsection*{Depolarizing noise}
The Kraus operators for the depolarization are equal to
\begin{equation}\label{kdep}
K^c_0=\sqrt{1-\frac{3\lambda_c}{4}}I;\;K^c_i=\sqrt{\frac{\lambda_c}{4}}\sigma_i
\end{equation}
where $I$ is the 2x2 identity matrix,  $\sigma_i$, $i=1,2,3$ are the Pauli matrices $X,Y,Z$, and $\lambda_c,\,c=A,B$ denotes depolarization strength for Alice and Bob respectively.

Depolarization channel is symmetric, i.e. it influences all bases in the same way and it can be intuitively understood as a uniform reduction of a Bloch ball radius. By introducing a parameter $V_c=1-\lambda_c$ one can express the probabilities of correct result and error in a similar form as for the dephasing channel 
\begin{equation}
 F_\phi=\frac{1+V_AV_B}{2};\;\; D_\phi=\frac{1-V_AV_B}{2}.
\end{equation}

\section{{PKE approximation}}\label{app:PKE}

{We derive an approximate formula for the PKE for the four-state BBM92 protocol in the regime of low noise parameter $n_b/\eta\ll 1$. 
We assume the same channels on Alice's and Bob's side, $\eta_A=\eta_B=\eta$, $n_A=n_B=n_b$.

First, note that one can assume that $\xi=2^mn_b/(2\eta p_{\textrm{pair}})$ is a small parameter, since otherwise the received signal strength would be similar to the noise gathered in the whole $2^m$ frame, which typically forbids secret key distillation. This is also confirmed by numerical calculations presented in Figs.~\ref{fig3} and~\ref{fig:PKE_pair}.

In order to find an approximate expression for the key rate one can expand Eq.~\eqref{bbm92} in the difference between QBERs in $X$ and $Z$ bases $e_X-e_Z\ll 1$, which is a valid approximation for high visibilities. One obtains 
\begin{equation}\label{eq:K_appr}
K^{\textrm{BBM92}}_{s=4}\approx q\Big[1-2h(e_{\textrm{av}})\Big]+\frac{(e_X-e_Z)^2}{4e_{\textrm{av}}(1-e_{\textrm{av}})\ln 2},
\end{equation}
where
$e_{\textrm{av}}={(e_X+e_Z})/{2}$ is the average QBER. Note that for the symmetric BBM92 protocol or depolarizing channel model $e_{\textrm{av}}=e_{\textrm{bit}}$. For the dephasing channel model and asymmetric protocol the correction to the key fraction is proportional to $(e_X-e_Z)^2/e_{\textrm{av}}\approx D_-^2/D$, where $D_-=D_X-D_Z$ and $D=(D_X+D_Z)/2$ are differential and average disturbances. We assume a high but nonideal visibility $0\neq D\ll 1$ and thus we will neglect this term in the main approximation.

One can now put the above formula Eq.~\eqref{eq:K_appr} into Eq.~\eqref{eq:PKE} and expand it in $\xi$ up to the first order resulting in
 \begin{align}\label{PKEapp}
     \textrm{PKE}^{\textrm{BBM92}}_{s=4}\approx qm\left\{(1+2p_{\textrm{pair}})\left[1-2h\left(\bar{e}_{\textrm{av}}\right)\right]+\right. \\ \left.
     +2p_{\textrm{pair}}\xi\log_2\left[4\bar{e}_{\textrm{av}}^2(1-\bar{e}_{\textrm{av}})^2\right]\right\} \notag 
 \end{align}
 where $\bar{e}_{\textrm{av}}=[p_{\textrm{pair}}+D(1-p_{\textrm{pair}})]/2$ is the average QBER for $\xi=0$ and we left only highest order terms in $p_{\textrm{pair}}$. Plugging $\xi=2^mn_b/(2\eta p_{\textrm{pair}})$ into Eq.~(\ref{PKEapp}) and maximizing with respect to coding order $m$ treated for simplicity as a continuous parameter one finds optimal number of qubits %$\bar{m}$ 
 \begin{equation}
     m^{*}=\frac{ W\left(\Xi\right)-1}{\ln 2},
 \end{equation}
 where $W(x)$ is the Lambert function and
 \begin{align}\label{eq:xi}
\Xi=-\frac{(1+2p_{\textrm{pair}})[1-2h(\bar{e}_{\textrm{av}})]}{\log_2[4\bar{e}_{\textrm{av}}^2(1-\bar{e}_{\textrm{av}})^2]}\frac{e\eta}{n_b}.
\end{align}

Note that since $\Xi\sim \eta/n_b \gg 1$ the Lambert function $W(\Xi)$ changes slowly so one can put $p_{\textrm{pair}}=0$ in Eq.~\eqref{eq:xi} without changing $m^{*}$ as long as the visibility is 
in the practical range $V<99.9\%$. 
One can therefore obtain
\begin{equation}\label{}
\Xi=-\frac{e\eta}{2n_b}\frac{1-2h\left(D\right)}{\log_2\left[2D\left(1-{D}\right)\right]}.
\end{equation}
To find the pair production probability one can minimize the $\textrm{QBER}=R_{\textrm{error}}/R_{\textrm{event}}$, resulting in
\begin{align}\label{eq:opt_p_app}
p_{\textrm{pair}}^{*}=\frac{1}{2}\left(\frac{\eta}{2n_b 2^{m^{*}}}-1\right)^{-1}=\frac{1}{2}\left(\frac{e\eta W(\Xi)}{2n_b\Xi}-1\right)^{-1}.
\end{align}
The final approximation for PKE is therefore given by
\begin{equation}
\textrm{PKE}^{\textrm{BBM92}}_{s=4}=q\left[1-2h\left(D+ \frac{{p}^{*}_{\textrm{pair}}}{2}\right)\right]\frac{[W(\Xi)-1]^2}{W(\Xi)\ln{2}}.
\end{equation}
and the corresponding QBERS can be obtained from expansion of $e_i=R_{\textrm{error}}^{i}/R_{\textrm{event}}^{i}$ and $p_{\textrm{pair}}$ as
\begin{equation}
e^{*}_{{i}}=\frac{p^{*}_{\textrm{pair}}}{2}+D_i\left(1-p^{*}_{\textrm{pair}}\right)+(1-3p^{*}_{\textrm{pair}})\frac{2^{m^*}n_b}{\eta}(1-2D_i),
\end{equation}
for $i=X,Z$, and $R_{\textrm{error}}^i,\,R_{\textrm{event}}^i$ denote error and event rates for the $i$-th basis.

\bibliography{multiqubit}

\end{document}